\documentclass[prl,aps,twocolumn,superscriptaddress,showpacs]{revtex4}
\usepackage{epsfig}
\usepackage{amsmath}
\usepackage{amsfonts}
\usepackage{amssymb,mathrsfs}
\usepackage{bm}
\usepackage{bbm}

\newcommand{\s}{\sum\limits}

\newcommand{\pa}{\partial}

\newcommand{\be}{\begin{equation}}
\newcommand{\e}{\end{equation}}
\newcommand{\beml}{\begin{subequations}}
\newcommand{\eml}{\end{subequations}}
\newcommand{\beq}{\begin{eqnarray}}
\newcommand{\eq}{\end{eqnarray}}
\newcommand{\ba}{\begin{array}}
\newcommand{\ea}{\end{array}}
\newcommand{\bpm}{\begin{pmatrix}}
\newcommand{\epm}{\end{pmatrix}}
\newcommand{\bc}{\begin{cases}}
\newcommand{\ec}{\end{cases}}
\newcommand{\lt}{\left}
\newcommand{\rt}{\right}

\newcommand{\la}{\langle}
\newcommand{\ra}{\rangle}
\newcommand{\ep}{\varepsilon}

\newcommand{\bb}{\boldsymbol}

\DeclareMathOperator{\sign}{sign}

\begin{document}

\title{Giant magneto-drag in graphene at charge neutrality}

\author{M. Titov}
\affiliation{Radboud University Nijmegen,
  Institute for Molecules and Materials, NL-6525 AJ Nijmegen, The
  Netherlands}

\author{R. V. Gorbachev} \affiliation{School of Physics and Astronomy,
  University of Manchester, Manchester M13 9PL, UK}
\affiliation{Centre for Mesoscience and Nanotechnology, University of
  Manchester, Manchester M13 9PL, UK}

\author{B. N. Narozhny} \affiliation{Institut f\"ur Theorie der
  Kondensierten Materie and DFG Center for Functional Nanostructures,
  Karlsruher Institut f\"ur Technologie, 76128 Karlsruhe, Germany}

\author{T. Tudorovskiy} \affiliation{Radboud University Nijmegen,
  Institute for Molecules and Materials, NL-6525 AJ Nijmegen, The
  Netherlands}

\author{M. Sch\"utt} \affiliation{Institut f\"ur Nanotechnologie,
  Karlsruhe Institute of Technology, 76021 Karlsruhe, Germany}

\author{P. M. Ostrovsky} \affiliation{Max-Planck-Institut f\"ur
  Festk\"orperforschung, Heisenbergstr. 1, 70569, Stuttgart, Germany}
\affiliation{Institut f\"ur Nanotechnologie, Karlsruhe Institute of
  Technology, 76021 Karlsruhe, Germany} \affiliation{L.D. Landau
  Institute for Theoretical Physics RAS, 119334 Moscow, Russia}

\author{I. V. Gornyi} \affiliation{Institut f\"ur Nanotechnologie,
  Karlsruhe Institute of Technology, 76021 Karlsruhe, Germany}
\affiliation{A.F. Ioffe Physico-Technical Institute, 194021
  St. Petersburg, Russia}

\author{A. D. Mirlin} \affiliation{Institut f\"ur Nanotechnologie,
  Karlsruhe Institute of Technology, 76021 Karlsruhe, Germany}
\affiliation{Institut f\"ur Theorie der Kondensierten Materie and DFG
  Center for Functional Nanostructures, Karlsruher Institut f\"ur
  Technologie, 76128 Karlsruhe, Germany} \affiliation{Petersburg
  Nuclear Physics Institute, 188350 St. Petersburg, Russia}

\author{M. I. Katsnelson} \affiliation{Radboud University Nijmegen,
  Institute for Molecules and Materials, NL-6525 AJ Nijmegen, The
  Netherlands}

\author{K. S. Novoselov} \affiliation{School of Physics and Astronomy,
  University of Manchester, Manchester M13 9PL, UK}

\author{A. K. Geim} \affiliation{School of Physics and Astronomy,
  University of Manchester, Manchester M13 9PL, UK}
\affiliation{Centre for Mesoscience and Nanotechnology, University of
  Manchester, Manchester M13 9PL, UK}

\author{L. A. Ponomarenko} \affiliation{School of Physics and
  Astronomy, University of Manchester, Manchester M13 9PL, UK}

\date{\today}

\begin{abstract}
 We report experimental data and theoretical analysis of Coulomb drag
 between two closely positioned graphene monolayers in weak magnetic
 field. Close enough to the neutrality point, coexistence of electrons
 and holes in each layer leads to a dramatic increase of the drag
 resistivity. Away from charge neutrality, we observe non-zero Hall
 drag. The observed phenomena are explained by decoupling of electric
 and quasiparticle currents which are orthogonal at charge
 neutrality. The sign of magneto-drag depends on the energy relaxation
 rate and geometry of the sample.
\end{abstract}


\maketitle

Recent measurements \cite{Gorbachev12} of frictional drag in
graphene-based double-layer devices revealed the unexpected phenomenon
of giant magneto-drag at the charge neutrality point. Applying
external magnetic fields as weak as 0.1-0.3 T results in the reversal
of the sign and a dramatic enhancement of the amplitude of the drag
resistance. If the device is doped away from charge neutrality, the
impact of such a weak field on the drag resistance is very modest. The
observed effect weakens at low temperatures, hinting at the classical
origin of the phenomenon.

In this Letter we report experimental data on longitudinal and Hall
drag resistivity in isolated graphene layers separated by a $1$\,nm
thick boron-nitride (hBN) spacer. The observed effects are explained
in terms of coexisting electron and hole liquids in each layer
\cite{Foster09,ryz}. This theory is based on the hydrodynamic
description of transport in graphene derived in
Refs.~\onlinecite{Mueller08,Schuett13,lux} using the quantum kinetic
equation framework \cite{Fritz08,Kashuba08}. It provides a simplified
description of the magneto-drag effect while capturing the essentially
classical physics of the phenomenon \cite{fn1}. The effect can be
traced back to the fact that the Lorentz force in the electronic band
is opposite to that in the hole band, which is also the reason for the
anomalously large Nernst effect \cite{Zuev09,Wei09} and vanishing Hall
effect at charge neutrality.

\begin{figure}[ht]
\centerline{\includegraphics[width=\columnwidth]{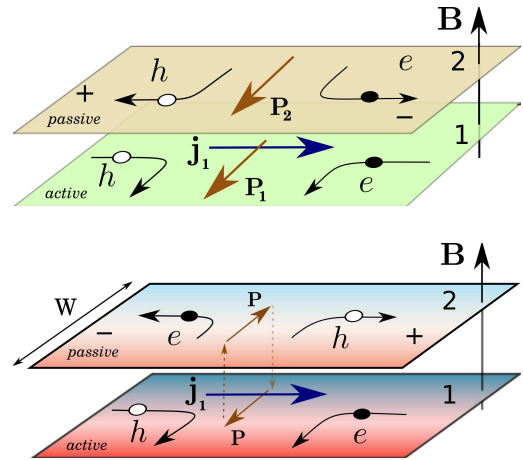}}
\caption{(Color online) Mechanism of magneto-drag at charge
  neutrality. Upper panel: in an infinite system quasi-particle
  currents in the two layers (denoted by ${\bb P}_i$) flow in the same
  direction, leading to {\em positive} drag
  $\rho^D_{xx}=V/j_1>0$. Lower panel: in a thermally isolated system
  no net quasi-particle flow is possible (leading to inhomogeneities
  in the quasi-particle density); the quasi-particle currents in the
  two layers have opposite directions yielding {\em negative} drag.}
\vspace*{-0.5cm}
\label{fig:setup1}
\end{figure}

The classical mechanism behind the giant magneto-drag is illustrated
in Fig.~\ref{fig:setup1}. The upper panel shows two \emph{infinite}
graphene layers at charge neutrality. The driving electric current
$\bb{j}_1$ in the active layer corresponds to the the
counter-propagating flow of electrons and holes with zero total
momentum due to exact electron-hole symmetry (hence, in the absence of
additional correlations there is no drag at the Dirac point
\cite{Schuett13,DSarma07,Narozhny12,Polini12,Peres12}). In a weak
magnetic field, electrons and holes are deflected by the Lorentz force
and drift in the same direction. The resulting quasi-particle flow,
$\bb{P}_1$, carries a non-zero net momentum in the direction
perpendicular to the charge flow, $\bb{j}_1$. The momentum transfer by
the interlayer Coulomb interaction induces the quasi-particle current
$\bb{P}_2$ in {\em the same} direction as $\bb{P}_1$. The Lorentz
forces acting on both types of carriers in the passive layer drive the
charge flow in the direction opposite to $\bb{j}_1$. If the passive
circuit is open, this current is compensated by a finite drag voltage $V$,
yielding a {\it positive} drag resistivity $\rho^D_{xx}=V/j_1>0$.

This mechanism of magneto-drag at charge neutrality is closely related
to the anomalous Nernst effect in single-layer graphene
\cite{Mueller08,Zuev09,Wei09}. Indeed, the quasi-particle current is
proportional to the heat current at the Dirac point. A similar
mechanism, where the role of $\bb{P}_i$ is played by a spin current,
has been proposed in Ref.~\cite{Abanin11} as a possible explanation
for a giant non-local magnetoresistance at charge neutrality.

The above argument describes the steady state in the infinite system
where all physical quantities are homogeneous in real space. This is
not the case in a relatively small mesoscopic sample. Whether a
particular sample should be considered ``small" or ``large" is
determined by comparing the sample size to the typical length scale
corresponding to the leading relaxation process. At high enough
temperatures, the heat currents are most efficiently relaxed by
electron-phonon scattering, which we describe in this Letter by the
length scale $\ell_{\textrm{ph}}$ \cite{Levitov12,fn2}.

In a finite system, the quasi-particle currents must vanish at the
boundaries.  For $W\gg \ell_{\textrm{ph}}$, the quasi-particle current
and density is homogeneous in the bulk and the system remains
effectively infinite.

For $W\ll \ell_{\textrm{ph}}$, the currents $\bb{P}_i$ acquire a
dependence on the coordinate $y$. In this case energy conservation
dictates that $\bb{P}_2(y)=-\bb{P}_1(y)$. As the result the electric
charge in the passive layer tends to flow \emph{in the same direction}
as $\bb{j}_1$, see Fig.~\ref{fig:setup1}. Thus, the
drag is {\it negative}, which is the conventional sign for the Coulomb
drag in a system with the same type of charge carriers.

In order to test the above ideas, we perform new measurements of the
drag effect in magnetic field which are illustrated in
Fig.~\ref{fig:results}. The experiments \cite{sup} are carried out on
graphene double-layer structure with 1 nm hBN spacer and two
electrostatic gates. The schematics of the experiment is shown in the
inset of the panel D in Fig.~\ref{fig:results}. The same device was
used in Ref.~\onlinecite{Gorbachev12} for drag measurements in zero
magnetic field.

The map for the drag resistivity, $\rho^D_{xx}(V_T, V_B)$, is shown in
Fig.~\ref{fig:results}, panel A at $T=240$ K. The main difference
compared to zero field experiment reported earlier \cite{Gorbachev12}
is large negative drag at the double neutrality point. A dramatic
change in drag resistivity with the applied magnetic field is shown in
more details in Fig.~\ref{fig:results}, panels B and E (at $160$ K
and $240$ K, respectively. To ensure same charge densities $n_1$ and
$n_2$ in the top and bottom layers, we sweep both gates simultaneously
along the line connecting the bottom left and top right corners of the
map. The experiment shows a large negative drag resistivity close to
the double neutrality point, $n_1 = n_2 = 0$, as expected for a small
sample (see above); in our device both layers have the width $W\approx
2\mu$m and sufficiently resistive contacts.

In addition to the longitudinal drag resistivity we also measure the
Hall drag resistivity, $\rho^D_{xy}(V_T, V_B)$, shown in
Fig.~\ref{fig:results}, panel C at $T=240$ K as a function of the top
and bottom gate voltages. Due to the low density of states in graphene
and the small separation between layers (in this experiment $d\approx
1$ nm), the relationship between gate voltages and charge densities is
rather nontrivial. To identify sign of charge carriers at each point
in Fig.~\ref{fig:results}, panel C, we also measured resistivity maps
for both layers. Since the resistance of graphene is peaked at charge
neutrality, tracking the position of the resistivity maximum gives the
lines which divide the map into the electron- and hole-doped
parts. Such lines are shown in both maps, see
Fig.~\ref{fig:results}, panels C and A. The observed Hall drag
resistance is large when one of the layers is close to neutrality
point and vanishes if two layers have the same charge densities with
opposite signs (a white line running from the top left to bottom right corner).

\setlength{\unitlength}{\columnwidth}
\begin{figure*}[t]
\centerline{\includegraphics[width=2\columnwidth]{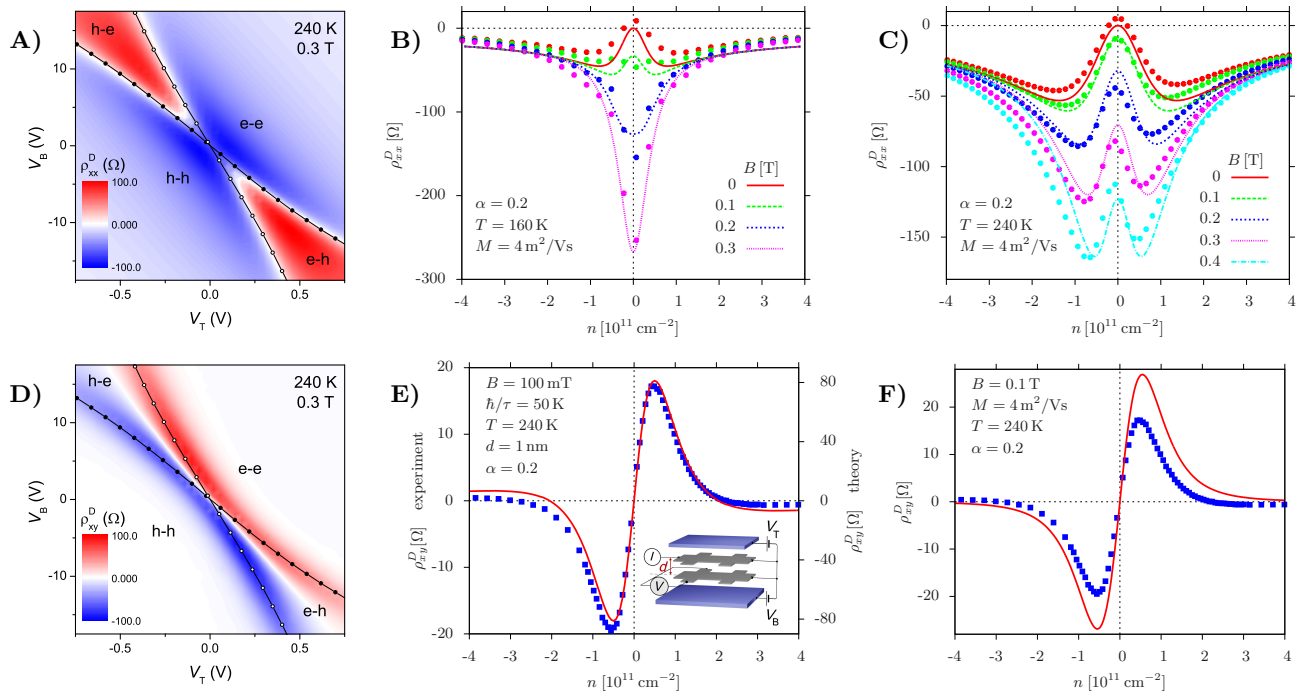}}
\caption{(Color online) Panel A: Longitudinal drag resistivity in
  magnetic field as a function of the top and bottom gate
  voltages. Lines track positions of the maxima in single-layer
  resistivity in top (open symbols) and bottom (closed symbols)
  layers. Panel B: Magneto-drag for equal charge densities $n_1=n_2=n$
  at $T=160$ K. Solid symbols represent the experimental data. The
  lines show the results of the proposed theory (\ref{Model2}), see
  Supplemental Material \cite{sup} for details. Panel C: The map of
  Hall drag resistivity as function of top and bottom gate voltages.
  The white diagonal corresponds to vanishing Hall drag for
  $n_1=-n_2$. The lines are similar to those in the Panel A. Panel D:
  The experimental data (blue squares, left axis) and theoretical plot
  (red solid line, right axis) for the Hall drag resistivity for equal
  charge densities $n_1=n_2=n$. The theoretical curve is calculated on
  the basis of the microscopic theory of Ref.~\onlinecite{Schuett13}.
  Note the sign change at $n\approx\pm 2\times 10^{11}$cm$^{-2}$.
  Inset: schematics of Hall drag measurements in double layer
  system. The charge density is controlled by voltages $V_T$ and $V_B$
  applied to the top and bottom gate, respectively. Panel E:
  Magneto-drag for equal charge densities $n_1=n_2=n$ at $T=240$
  K. Solid symbols represent the experimental data. The lines show the
  results of the proposed theory (\ref{Model2}). Panel F: The
  experimental data (blue squares) and the results of the proposed
  theory (\ref{Model2}) (red solid line) for the Hall drag resistivity
  for equal charge densities $n_1=n_2=n$. The data are identical to
  those in Panel D.}
\vspace*{-0.3cm}
\label{fig:results}
\end{figure*}

We now turn to the theoretical description of the drag
effect. Consider first the Drude model for electrons and holes in two layers,
\beml
\label{Drude0}
\beq
\label{1}
e\bb{E}_i+e[\bb{v}_{ie}\times\bb{B}] &=& \bb{F}_{ie} + e\, \bb{v}_{ie}/M_i,\\
\label{2}
-e\bb{E}_i-e[\bb{v}_{ih}\times\bb{B}] &=& \bb{F}_{ih} + e \, \bb{v}_{ih}/M_i,
\eq 
\eml 
where $i=1,2$ and $\bb{v}_{ia}$, $a=e,h$, stand for the mean
velocities of electrons and holes in the layer $i$, $\bb{E}_{1,2}$ and
$\bb{B}$ are the electric and magnetic fields, $e$ is the elementary
charge, and $M_{1,2}$ are the carrier mobilities due to scattering on
impurities. The electric $\bb{j}_i$ and quasi-particle $\bb{P}_i$
currents are related to $\bb{v}_{ia}$ by \cite{Foster09}
\be
\label{currents}
\bb{j}_i=e( n_{ie} \bb{v}_{ie} - n_{ih} \bb{v}_{ih}), \qquad 
\bb{P}_i=n_{ie} \bb{v}_{ie} + n_{ih} \bb{v}_{ih},
\e
with $n_{ie(h)}=\int_0^\infty d\ep\,
\nu(\ep)\lt[e^{(\ep\mp\mu_i)/T}+1\rt]^{-1}$ standing for the electron
and hole densities. Here $\nu(\ep)=2 |\ep|/\pi(\hbar v)^2$ is the
density of states in graphene (disregarding the magnetic field), and
$\mu_i$ are the chemical potentials measured from the Dirac point. The
total charge and quasi-particle densities are defined as
$n_i=n_{ie}-n_{ih}$ and $\rho_i=n_{ie}+n_{ih}$.

In general, the frictional force acting on each type of carriers can
be represented by the sum
\be
\label{friction}
\bb{F}_{ia} = \hbar\s_{jb} \lt[\gamma^{ab}_{ij} n_{jb} (\bb{v}_{ia}\!-
  \bb{v}_{jb}) +\tilde{\gamma}^{ab}_{ij} n_{jb} (\bb{v}_{ia}\!+
  \bb{v}_{jb})\rt],
\e
where the coefficients $\tilde{\gamma}$ appear in monolayer graphene
due to the absence of Gallilean invariance. The expression
(\ref{friction}) can be obtained by solving the quantum kinetic
equation (QKE) in the hydrodynamic approximation \cite{us2}. The
dimensionless coefficients $\gamma^{ab}_{ij}$ and
$\tilde{\gamma}^{ab}_{ij}$ are related to microscopic collision rates
\cite{Schuett13,lux,Mueller08,Fritz08}.

For $n_i=0$, the first term in Eq.~(\ref{friction}) simplifies to
\be
\label{sim}
\bb{F}_{1a}=-\bb{F}_{2a} = \hbar\, \gamma\, (\bb{P}_1-\bb{P}_2), 
\e
where $\gamma = \hbar/ T\tau_P$, with $\tau_P^{-1}$ being
the momentum relaxation rate. The second term in Eq.~(\ref{friction})
renormalizes the mobilities \cite{Schuett13,Fritz08,Kashuba08}. The Drude
model (\ref{Drude0}) with the force (\ref{sim}) also describes the
case $\mu_1=\mu_2 \gg T$, where $\gamma = \hbar/ \mu_1\tau_P$. In both
limits, the model (\ref{Drude0}) is equivalent to the
hydrodynamic transport equations derived from the QKE
\cite{Schuett13,us2}.

For strongly-doped graphene, $\mu_i\gg T$, the quasi-particle current
and density are obsolete: $\bb{P}_i=\bb{j}_i/e$ and $\rho_i=n_i$.
Equations (\ref{Drude0}) are then reduced to the standard Drude
model yielding the vanishing Hall drag resistivity
$\rho_{xy}^D=E_{2y}/j_{1x}=0$ and conventional drag,
$\rho^D_{xx}=E_{2x}/j_{1x}=-\hbar \gamma/e^2$, which show negligible
dependence on the magnetic field \cite{Jauho93}.

In contrast, at charge neutrality the quasi-particle and charge
degrees of freedom are decoupled. The quasi-particle density for
$n_i=0$ is determined by the temperature as $\rho_i=\rho_0=\pi
T^2/3(\hbar v)^2$, while the electric and quasi-particle currents
$\bb{j}_1$ and $\bb{P}_i$ become orthogonal, as shown in
Fig.~\ref{fig:setup1}.
 
Rewriting Eqs.~(\ref{Drude0}), (\ref{sim}) in terms of currents and
excluding $\bb{P}_i$, we obtain the resistivity tensor. At charge
neutrality, $n_i=0$, the longitudinal drag resistivity is peaked and
its value is given by the expression
\be
\label{res1}
\rho_{xx}^D= \frac{\hbar\, \gamma}{e^2}\, \frac{B^2
  M_1M_2}{1+\hbar\,\gamma\, \rho_0 (M_1+M_2)/e},\qquad n_i=0, 
\e 
which describes positive drag in an infinite system in agreement with
the qualitative picture illustrated in Fig.~\ref{fig:setup1}, upper
panel. In the limit of weak interaction, $\gamma M T^2 \ll \hbar e
v^2$, the result (\ref{res1}) can be obtained from the standard
perturbative approach \cite{Narozhny12} modified for graphene in a
classical magnetic field.

Large negative peak in $\rho_{xx}^D$ at the double neutrality point
(Fig.~\ref{fig:results} panel B) suggests that the sample width,
$W\approx 2\,\mu$m is relatively small as compared to
$\ell_{\textrm{ph}}$ (Fig.~\ref{fig:setup1}, lower panel). To account
for the finite sample width, we re-write the equations (\ref{Drude0})
in terms of the currents $\bb{j}_i$ and $\bb{P}_i$ and allow for the
spatially varying quasi-particle density, $\rho_i$, in the sample.
The resulting model for the first layer reads
\beml
\label{Model2}
\beq
\label{enE}
- K_1 \bb{\nabla} \rho_1+e n_1\bb{E}_1 +[\bb{j}_1\times \bb{B}] 
&=& 
\rho_1 \bb{F}_{1} + e\bb{P}_1/M_1,
\mbox{\quad  }
\\
\label{erhoE}
e \rho_1\bb{E}_1 + e[\bb{P}_1\times \bb{B}] &=& n_1 \bb{F}_{1} +  \bb{j}_1/M_1,
\\
\label{continuity}
\bb{\nabla P}_1 = -( \rho_1-\rho_0)/\tau_{\textrm{ph}}
&-&(\rho_1-\rho_2)/(2\tau_Q).  
\eq 
\eml 
One has to replace the index $1$ with $2$ for the second layer.
Here $K_1=(\pi \hbar^2 v^2/2) \pa n_1/\pa\mu=2T\ln (2 \cosh \mu_1/2T)$ is the
mean quasi-particle kinetic energy. The continuity equation for the
quasi-particle current (\ref{continuity}) includes relaxation by the
electron-hole recombination \cite{Foster09}, with
$\tau_{\textrm{ph}}^{-1}$ describing the energy loss from the system,
which at high enough temperatures is dominated by phonon scattering
\cite{fn2}, and $\tau_{Q}^{-1}$ describing the quasiparticle imbalance
relaxation due to the interlayer Coulomb interaction. For
$\tau_{\textrm{ph}}^{-1}=0$, one finds $\bb{P}_1+\bb{P}_2=0$ because
the interlayer interaction does not lead to relaxation of the total
quasiparticle current. Near the Dirac point, the energy and momentum
relaxation rates coincide (in particular, $\tau_{Q}\sim \tau_{P}$). In
doped graphene, the recombination rates are exponentially suppressed
\cite{sup}.

The continuity equation for the electric current simply reads
$\bb{\nabla} \bb{j}_i =0$, hence $\bb{j}_i=(j_i(y),0)$. Within linear
response, the density $\rho_i$ has to be substituted by the
equilibrium value $\rho_0$ in the products $\rho_i \bb{F}$ and
$\rho_i\bb{E}$. This way we obtain the linear system of differential
equations on the functions $P_{iy}(y)$, $j_{1x}(y)$, and
$\rho_i(y)$. Since the charge current acquires the dependence on the
transverse coordinate, we define $\rho_{xx}^D=E_{2x}/\la j_{1x}\ra$,
where $\la j_{1x}\ra=W^{-1}\int_0^W j_{1x}dy$.

The model (\ref{Model2}) with the frictional force (\ref{sim}) admits
a full analytic solution \cite{sup} in terms of the relaxation rates
$\tau_Q^{-1}$, $\tau_{\rm ph}^{-1}$, and $\tau_P^{-1}$. The resulting
behavior crucially depends on these rates: in particular, in the
absence of phonons ($\tau_{\rm ph}\rightarrow\infty$, i.e. in a
thermally isolated system) the drag at the Dirac point is always
negative, in agreement with the qualitative picture of
Fig.~\ref{fig:setup1}. For vanishing sample width ($W\rightarrow 0$)
we find $\rho_{xx}^D\approx - B^2W^2/(24\rho_0 K \tau_Q)$. In general,
these rates depend on the carrier density and have to be determined by
the microscopic theory \cite{Schuett13}. Relegating further
mathematical details to the online Supplemental Material \cite{sup},
we present the results of the model calculations in
Fig.~\ref{fig:results} alongside experimental data.

The drag resistivity $\rho_{xx}^D$ is plotted in
Fig.~\ref{fig:results}, panels B and E for $T=160$\,K and $T=240$\,K,
respectively. The collapse of the theoretical curves at high carrier
density is an artifact of the phenomenological model \cite{fn2}, which
is most reliable near charge neutrality. At higher temperature
(Fig.~\ref{fig:results}, panel E), the drag resistivity exhibits
qualitatively new features near charge neutrality which can be
physically attributed to higher efficiency of relaxation
processes. The sign of $\rho_{xx}^D$ at the Dirac point is then
determined by the relation between the typical relaxation length
$\ell_{\rm ph}= 2\sqrt{K\tau_{\rm ph} M /e}$ and the sample
width. This is illustrated in Fig.~\ref{fig:Wdep}, where we plot
$\rho_{xx}^D$ as a function of magnetic field for different values of
the sample width choosing realistic values for the model parameters
$T=240$\,K, $M=4$\,m$^2/$Vs, and $\ell_\textrm{ph} =1.2\,\mu$m.

\begin{figure}[t]
\centerline{\includegraphics[width=0.9\columnwidth]{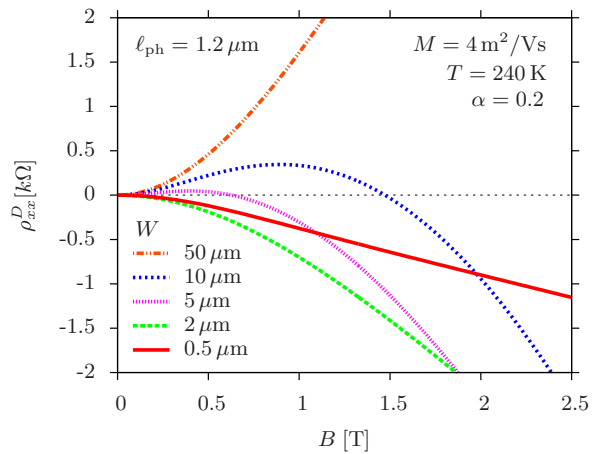}}
\caption{(Color online) The magnetic field dependence of the
  longitudinal drag resistivity at the neutrality point. The positive
  sign of the magneto-drag in weak fields corresponds to the limit
  $W\gg \ell_{\textrm{ph}}$, where $\ell_\textrm{ph} \approx 1.2\,\mu$m
  for the parameters of the plot. The magnetic field dependence of
  scattering rates is disregarded in the plot.}
\vspace*{-0.3cm}
\label{fig:Wdep}
\end{figure}

Based on the above results, we predict that in wider samples, the
giant magneto-drag at the Dirac point should become positive. We also
speculate that the magneto-drag at the double Dirac point may become
positive in stronger fields due to the magnetic-field dependence of
the scattering times $\tau_\textrm{Q}$, $\tau_\textrm{P}$, and
$\tau_{\textrm{ph}}$.

The model (\ref{Model2}) allows us to calculate the Hall drag
resistivity $\rho_{xy}^D$. The result is shown in
Fig.~\ref{fig:results}, panel F. The theory also predicts the
vanishing Hall drag for the case of oppositely doped layers,
$n_1=-n_2$. Interestingly enough, the data shows a sign change of
$\rho_{xy}^D$ at $n\approx\pm 2\times 10^{11}$cm$^{-2}$. At that point
the effect is rather weak and requires a more accurate
consideration. Using the microscopic theory of
Ref.~\onlinecite{Schuett13}, we have evaluated the Hall drag
resistivity for an infinite sample with an energy-independent impurity
scattering time $\tau$. The value of $\tau$ was determined from the
single-layer resistivity measured in experiment and we have used the
most plausible estimate for the value of the effective
electron-electron interaction strength, $\alpha\approx 0.2$, in
graphene on hBN. The result is shown in Fig.~\ref{fig:results}, panel
D along with the corresponding experimental data without any fitting.

In conclusion, we have measured the longitudinal and Hall drag
resistivity in double-layer graphene and provided a theoretical
description of the observed effects. The giant magneto-drag at the
neutrality point appears due to the presence of two types of carriers
(electrons and holes), which in weak magnetic fields experience a
unidirectional drift orthogonal to the driving current. This effect is
specific to the neutrality point, where non-zero drag appears despite
the exact electron-hole symmetry. The present theory does not rely on
the Dirac spectrum in graphene, but is equivalent to the microscopic
theory \cite{Schuett13,fn1} at and far away from the charge neutrality
thus capturing the essential physics of magneto-drag. For a more
accurate description of the effect at intermediate densities, the
microscopic theory should be formulated on the basis of the kinetic
equation \cite{us2}.

We are grateful to the Royal Society, the K\"orber Foundation, the
Office of Naval Research, the Air Force Office of Scientific Research,
the Engineering and Physical Sciences Research Council (UK), Stichting
voor Fundamenteel Onderzoek der Materie (FOM, the Netherlands), DFG
SPP 1459 and BMBF for support.

Upon completion of this manuscript, we became aware of a related work
by Song and Levitov \cite{LevitovArXiv}.

\newpage
\onecolumngrid
\setcounter{enumiv}{0} 
\setcounter{equation}{0} 

\vspace{1cm}
\centerline{\bfseries ONLINE SUPPORTING INFORMATION}

\maketitle

\section{Experimental Devices and Measurement Procedures}
\label{sec:expDevices}

The double layer graphene devices were fabricated by using procedures
previously described in detail in Refs.~\onlinecite{Ponomarenko11} and
\onlinecite{Gorbachev12A}. The maximum size of our double-layer Hall
bars is currently limited to, typically, 2\,$\mu$m $\times$ 10\,$\mu$m
because of the formation of pockets of a hydrocarbon residue at the
interface between graphene and BN (for details, see Supplementary
Material in Ref.~\onlinecite{Ponomarenko11}). These pockets (or
bubbles) appear randomly, and our Hall bars are made to fit inside
clean patches between bubbles. This restricts the width of
double-layer devices to 1.5-2\,$\mu$m. Making smaller and narrower
Hall bars is impractical because of reduced mobility and difficulties
associated with alignment of two Hall bars exactly on top of each
other. Therefore, at present it is impossible to study magnetodrag in
devices of different widths. The reported Hall bars had a width of 2
$\mu$m, and the overlapping area for top and bottom Hall bars was
$\approx 15$\,$\mu$m$^2$.

All experimental results presented in this work were obtained by using
DC measurements with current commutation at each data point. We have
chosen DC over AC measurements because in the latter case a
significant out-of-phase signal (up to ~30\%) appears near the
neutrality point even at frequencies as lows as $\sim$ 10-30\,Hz. This
signal originates probably from capacitance coupling between the
closely spaced graphene layers. Each data point was measured for 12
seconds (6 sec for each polarity of the current), which corresponds to
an effective frequency of 0.2 Hz, low enough to avoid the capacitive
coupling. Further increase in the acquisition time improved accuracy
but did not affect the reported curves.

To ensure that the measurements are done in the linear response
regime, applied current was kept low. To this end, I-V curves were
measured at representative points of the reported maps. . We have
found empirically that a current of 50 nA can be used near the
neutrality point where the nonlinearity is largest. At higher carrier
densities, it is possible to increase the current by a factor of
10. The main reason for the nonlinearity is a voltage drop along the
Hall bar. This voltage can be comparable to the difference between
Fermi levels in the two graphene layers and, if the current is high,
the voltage drop can result in a gradient of carrier concentration
along the direction of current.

Furthermore, there is a finite tunnelling resistance through 3 atomic
layers of BN separating top and bottom graphene Hall bars. At low
biases used in all our measurements, the interlayer resistance $R_T$
exceeded 300\,k$\Omega$ over the whole range of gate voltages. This
value translates into a small contribution to the measured drag
resistivity estimated as \cite{Halperin}
\be
\label{estimate}
\delta \rho_{xx}^D = \frac{L}{W} \frac{\rho^2}{12 R_T}
\e
for the case of equivalent layers. At the neutrality point at $T=240$\,K we have 
$R_T = 500$\, k$\Omega$,  $\rho=1$\,k$\Omega$ is the intralayer resistivity (resistance per square)
and the aspect ratio $L/W\approx 2$ for our sample. This yields the insignificant contribution
$\delta \rho_{xx}^D \approx 0.33$\,$\Omega$.  

However, to ensure that the direct electrical coupling
does not affect the measured drag resistance, we have also compared
measurements with the passive graphene layer (the one where we measure
drag voltage) being floated and grounded. In latter case, several
different contacts were connected to the ground. If the tunnelling
current is significant the grounded contacts should act as a sink and
affect drag measurements. The difference was found insignificant for
trilayer BN devices, as expected from the estimate (\ref{estimate}). 
In contrast, devices with bilayer BN as a spacer exhibited very
significant changes indicating that bilayers are too transparent to be
used as an insulating layer for drag measurements.

Another important consideration in the reported measurements was to
ensure that there was no additional AC current flowing through the
devices due to external radiation at high frequencies, which could
increase electron temperature in graphene with respect to cryostat
temperature. To this end, radio frequency filters were fitted. Their
efficiency was judged by the observation of strong suppression of a
rectification signal (voltage in the absence of applied current),
which could be comparable to drag voltage if no filters were used.

\section{Thickness of BN determined from transport measurements}
\label{sec:expBN}

The thickness of BN separating graphene layers was routinely
determined during the device fabrication. To this end, we could use
several techniques including atomic force microscopy (AFM) \cite{Dean10}, optical contrast \cite{Gorbachev11,Golla13}, Raman spectroscopy \cite{Gorbachev11} and tunnelling \cite{Britnell12}. In
practice, the first two were found sufficient to determine the
thickness with single-layer accuracy. The tunnelling resistance for
the known device area was then used as an additional cross-check of the
BN thickness. In this section we show that transport measurements
provide yet another way to measure the thickness of the BN layer.

The approach is based on converting the Hall resistance measured in
weak magnetic fields into carrier concentration and then finding the
thickness as a fitting parameter from the gate voltage dependence of
carrier concentration. In general the density is nonlinear complicated
function of both gate voltages. However, if one of the layers is kept
at the neutrality point, the density in the other layer acquires a
relatively simple dependence on gate voltage. Our approach is similar
to the one recently used by Kim et al. \cite{Kim12}.

\begin{figure}
\centerline{
\includegraphics[width=0.5\columnwidth]{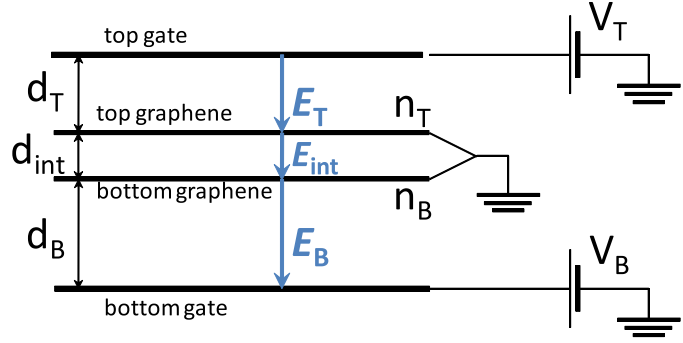}
}
\caption{Sketch for 4-plate capacitor model.}
 \label{fig:Set1}
\end{figure}

Consider a 4-plate capacitor in Fig.~\ref{fig:Set1}, which consists of
two grounded graphene sheets separated by a BN layer with thickness
$d_\textrm{int}$ and two gates, top and bottom, isolated by relatively thick BN,
with thicknesses $d_T$ and $d_B$, respectively. The charge densities
in both graphene layers $n_T$ and $n_B$ are related to the applied
gate voltages $V_T$ and $V_B$ through the system of nonlinear
equations:
\beq
&&
E_T\frac{d_T}{\ep}+\frac{1}{e}E_F(n_T)=V_T,
\label{ex1}
\\
&&
E_\textrm{int}\frac{d_\textrm{int}}{\ep}+\frac{1}{e}\lt(E_F(n_T)-E_F(n_B)\rt)
=0,
\label{ex2}
\\
&&
E_B\frac{d_B}{\ep}+\frac{1}{e}E_F(n_B)=-V_B,
\label{ex3}
\\
&&
E_T-E_\textrm{int}=\frac{en_T}{\ep_0},
\label{ex4}
\\
&&
E_\textrm{int}-E_B=\frac{en_B}{\ep_0},
\label{ex5}
\eq
where $E_T$, $E_B$ and $E_\textrm{int}$ are the electric fields in the
top, bottom and middle BN layers, respectively, $\varepsilon = 3.2$ is
the dielectric constant of BN \cite{Kim12}, $\varepsilon_0$ the
electric constant, $e$ the charge of electron and
$E_F(n)=\sign(n)\hbar v_F \sqrt{\pi |n|}$ the Fermi energy in graphene
($v_F \approx 10^6$\,m$/$s is the Fermi velocity; for simplicity we
assume $T= 0$ and no external doping). The Fermi energy is positive
for electrons ($n>0$ ) and negative for holes
($n<0$). Eqs. (\ref{ex1}-\ref{ex3}) describe the potential drop across
the top, middle and bottom BN, whereas Eqs. (\ref{ex4}-\ref{ex5})
follow from the Gauss law for the top and bottom graphene layers.

If one of the graphene layers is at the neutrality point (e.g. for
$n_T=0$), Eqs. (\ref{ex1}-\ref{ex5}) reduce to a one-line expression
that links gate voltage directly to carrier density in the
corresponding layer:
\be
\label{ex6}
V_B=\frac{qd_B}{\ep\ep_0}n_B+\sign(n_B)
\frac{\hbar v_F\sqrt{\pi |n_B|}}{q}\lt(1+\frac{d_B}{d_\textrm{int}}\rt).
\e
The first (linear) term gives the slope of $V_B(n_B)$ at high
densities and depends only on the thickness of the insulator
separating graphene from the gate (this term is just the classical
capacitance). The second (root-squared) term comes from the quantum
capacitance of graphene and is responsible for nonlinear behaviour of
the double-layer system close to the NP. The term depends on ratio
$d_B/d_\textrm{int}$.

\begin{figure}
\centerline{
\includegraphics[width=0.8\columnwidth]{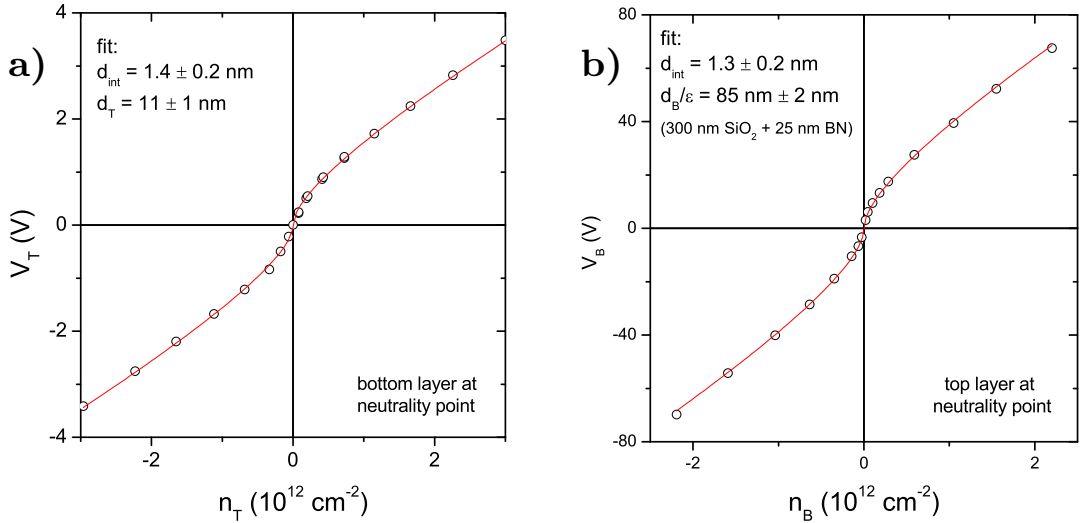}
}
\caption{BN thickness extracted from transport measurements. (a)
  Density in the top graphene layer as a function of gate voltage
  applied to this layer, provided the bottom graphene layer is kept
  neutral.  Open symbols -- experiment; solid curve -- best fit to
  Eq.~(\ref{ex7}). Different points correspond to different voltages
  applied to the bottom gate. (b) Same type of measurements as in (a)
  but with top and bottom layers interchanged. Both (a) and (b) give
  the same value of $d_\textrm{int}$.}
 \label{fig:exp2}
\end{figure}

Similar expression can be derived for  $n_B=0$:
\be
\label{ex7}
V_T=\frac{qd_T}{\ep\ep_0}n_T+
\sign(n_T)\frac{\hbar v_F\sqrt{\pi |n_T|}}{q}\lt(1+\frac{d_T}{d_\textrm{int}}\rt).
\e
Therefore the thickness $d_\textrm{int}$ can be determined from two
independent sets of measurements, presented in
Figs.~\ref{fig:exp2}a and \ref{fig:exp2}b. The experiment is carried
out as the following. We fix $V_B$ and ramp $V_T$ recording Hall
resistance for both layers. When the bottom layer reaches its
neutrality point we record values of $V_T$ and $n_T$. This gives one
data point in Fig.~\ref{fig:exp2}a. The same procedure is repeated for
several $V_B$ until enough data are collected. Then the same procedure
is carried out for fixed $V_T$, that is, the top and bottom layers are
effectively swapped in these measurements. Within experimental
accuracy, Figs.~\ref{fig:exp2}a and \ref{fig:exp2}b yield the same
value of $d_\textrm{int}$ which is $\approx 4$ times larger than interlayer
spacing for bulk graphite and BN. This agrees well with the expected
separation for the top and bottom graphene sheets if three BN layers
are in between. This value is also in agreement with the AFM
measurements carried on the same device.

\section{Phenomenological theory of double-layer devices}
\label{sec:theory}

Here we present the full analytic solution of the theoretical
model given by Eqs.~(6) and (4) of the main text.

\subsection{Model equations within linear response}

Under the assumptions of linear response we have to re-write the
equations~(6) of the main text as follows [the frictional force is taken in the form (4) of the main text; for brevity we write
equations for one layer only]
\begin{subequations}
\label{phm}
\begin{eqnarray}
\label{qeq}
&&
- \frac{\kappa_1}{en_1} \bb{\nabla} \rho_1+\bb{E}_1 + R_H^{(1)}[\bb{j}_1\times \bb{e}_z] =
e \frac{\rho_0}{n_1} \Gamma (\bb{P}_{1}-\bb{P}_2) + e R_0^{(1)}\bb{P}_1,
\\
&&
\nonumber\\
\label{jeq}
&&
\frac{\rho_0}{n_1}\bb{E}_1 + e R_H^{(1)}[\bb{P}_1\times \bb{e}_z]
= e \Gamma (\bb{P}_{1}-\bb{P}_2) + R_0^{(1)} \bb{j}_1
\\
&&
\nonumber\\
\label{ceq}
&&
\bb{\nabla}\cdot\bb{P}_1 = -\frac{\rho_1-\rho_0}{\tau_{\textrm{ph}}}
-\frac{\rho_1-\rho_2}{2\tau_\textrm{Q}}.
\end{eqnarray}
\end{subequations}
Here $R_0=1/(enM)$ and $R_H=B/(en)$ are the Drude and Hall resistances
of the single-layer graphene far from the Dirac point; and
$\Gamma=\gamma/e^2$ is the drag resistance far from the Dirac point
and in zero field.

The above equations should be supplemented by the following boundary conditions:
the quasiparticle currents must vanish at the edge of the sample
\begin{subequations}
\label{bc}
\begin{equation}
\bb{P}_i(y=0) = \bb{P}_i(y=W)=0,
\end{equation}
and on average no electric current is allowed in the passive layer
\begin{equation}
\label{av}
\overline{\bb{j}_2} = \frac{1}{W} \int\limits_0^W\!\! dy\, \bb{j}_2 = 0.
\end{equation}
\end{subequations}
The model (\ref{phm}) with the boundary conditions (\ref{bc}) admits a
full analytic solution. The results simplify in the case of identical
layers. The corresponding solutions are given below.

The results for inequivalent layers are qualitatively similar to the
below solutions. We illustrate the dependence of magnetodrag in the
double neutrality point on magnetic field in Fig.~\ref{fig:B}b for the case of
different mobilities in the layers: $M=7$\, m$^2/Vs$ in the active
layer and $M=3$\, m$^2/Vs$ and in the passive layer.

\subsection{Drag between identical layers}

Given the nonuniform current flow, the drag resistivity is defined
as the ratio of the averaged induced voltage in the passive layer to
the driving current in the active layer
\begin{equation}
\label{rd-def}
\rho_{xx}^D = \frac{\overline{E_{2x}}}{\overline{j_{1x}}},
\end{equation}
where the averaging was defined in Eq.~(\ref{av}).

\subsubsection{General expression}

The resulting expression reads
\begin{equation}
\label{res-rd}
\rho_{xx}^D = \frac{r_0}{2}\left[
\frac{1}{1 - \left[1-\frac{n^2}{\rho_0^2}\right] \frac{R_H^2}{R_H^2+R_0^2} \bar{f}_+}
-
\frac{1}{1 -\frac{n^2}{\rho_0^2} \frac{2\Gamma}{2\Gamma +r_0}
- \left[1-\frac{n^2}{\rho_0^2}\right]
\frac{R_H^2}{R_H^2+R_0^2+\frac{\rho_0^2-n^2}{n^2}2\Gamma r_0} \bar{f}_-}
\right],
\end{equation}
where $r_0=1/(e\rho_0 M)$ is the residual resistance of single-layer
graphene at the Dirac point in the absence of magnetic field, and
\be
\bar{f}_\pm = 1 - \frac{\tanh(W/L_\pm)}{W/L_\pm},
\e
with
\be
L_+^{-2} = \frac{e^2 n}{4\kappa\tau_{\rm ph}} R_0 \left[1+\frac{R_H^2}{R_0^2}\right],
\quad
L_-^{-2} = \frac{e^2 n}{4\kappa} R_0
\left[ \frac{1}{\tau_{\rm ph}} + \frac{1}{\tau_Q} \right]
\left[\frac{2\Gamma}{r_0}\left(1-\frac{n^2}{\rho_0^2}\right) + 1+\frac{R_H^2}{R_0^2}\right].
\e
Note, that $R_H/R_0 = BM$.

\subsubsection{Drude limit}

Far away from the Dirac point (i.e. for $\mu\gg T$) only one band
contributes (such that $n=\rho_0$) and the result (\ref{res-rd}) simplifies
to the standard Drude form which is independent of magnetic field
\be
\rho_{xx}^D (\mu\gg T) = - \Gamma.
\e

\subsubsection{Zero field limit}

In the absence of magnetic field $R_H=0$ and the result (\ref{res-rd})
simplifies to
\begin{equation}
\label{zfrd}
\rho_{xx}^D (B=0) = \frac{r_0}{2}
\left[1 - \frac{1}{1-\frac{n^2}{\rho_0^2}\frac{2\Gamma}{2\Gamma+r_0}} \right].
\end{equation}
Setting $n=\rho_0$ we recover the above Drude result. On the other
hand at the Dirac point $n=0$ and drag vanishes
\be
\rho_{xx}^D (B=0; n=0) = 0.
\e

\subsubsection{Neutrality point}

At the Dirac point ($n=0$) in the presence of magnetic field we find
\be
\rho_{xx}^D (n=0) = \frac{r_0}{2}
\left[
\frac{1}{1-\bar{f}_+ \frac{(BM)^2}{1+(BM)^2}}
-
\frac{1}{1-\bar{f}_- \frac{(BM)^2}{1+(BM)^2+2\Gamma/r_0}}
\right].
\e
Clearly, this result vanishes in the absence of the field.

\subsubsection{Neutrality point in an infinite sample}

Consider now the above result in the limit of an infinite sample $W\gg
L_+$. (Note that $L_+ > L_-$ by definition.)  Then $\bar f_\pm = 1$
and we find (in agreement with Eq.~(5) of the main text)
\be
\rho_{xx}^D (n=0; W\rightarrow \infty) = \frac{r_0}{2} \left[
\frac{1}{1-\frac{(BM)^2}{1+(BM)^2}}
-
\frac{1}{1-\frac{(BM)^2}{1+(BM)^2+2\Gamma/r_0}}
\right] =
\frac{\Gamma r_0}{2\Gamma + r_0} (BM)^2 >0.
\e
The resulting drag is {\it positive}.

\subsubsection{Neutrality point in a narrow sample}

For completeness, let us now consider now a limit of a
vanishing width $W\ll L_-$. Here $\bar f_\pm \approx
W^2/(24L_\pm^2)$ and expanding in this small factor we find
\be
\rho_{xx}^D (n=0; W\rightarrow 0) \approx
- r_0(BM)^2 \frac{e W^2}{24 M \kappa \tau_Q} < 0.
\e
The resulting drag is {\it negative}, independent of phonon scattering
time and mobility (this can be seen by taking into account that the
resistance $r_0$ is inversely proportional to the mobility).

\subsection{Hall drag between identical layers}

The Hall drag resistance is defined as
\begin{equation}
\label{rh-def}
\rho_{xy}^D = \frac{\overline{E_{2y}}}{\overline{j_{1x}}},
\end{equation}
where the averaging was defined in Eq.~(\ref{av}).

The resulting expression for the Hall drag resistance is
\begin{equation}
\label{res-rh}
\rho_{xy}^D = \frac{r_0}{2}\frac{n}{\rho_0} BM \left[
\frac{1}{1 - \left[1-\frac{n^2}{\rho_0^2}\right] \frac{R_H^2}{R_H^2+R_0^2} \bar{f}_+}
-
\frac{\frac{r_0}{2\Gamma+r_0} - \left(1-\frac{n^2}{\rho_0^2}\right)
\frac{2\Gamma \bar{f}_-}{r_0\left[1+(BM)^2 +
            \left(1-\frac{n^2}{\rho_0^2}\right)\frac{2\Gamma}{r_0}\right]}
}
{1 -\frac{n^2}{\rho_0^2} \frac{2\Gamma}{2\Gamma +r_0}
- \left[1-\frac{n^2}{\rho_0^2}\right]
\frac{R_H^2}{R_H^2+R_0^2+\frac{\rho_0^2-n^2}{n^2}2\Gamma r_0} \bar{f}_-}
\right],
\end{equation}
where it is clear that Hall drag vanishes in the absence of the
magnetic field and at the Dirac point as it should. It is also clear
that Hall drag vanishes in the Drude limit where $n=\rho_0$.

\begin{figure}
\centerline{
\includegraphics[width=\columnwidth]{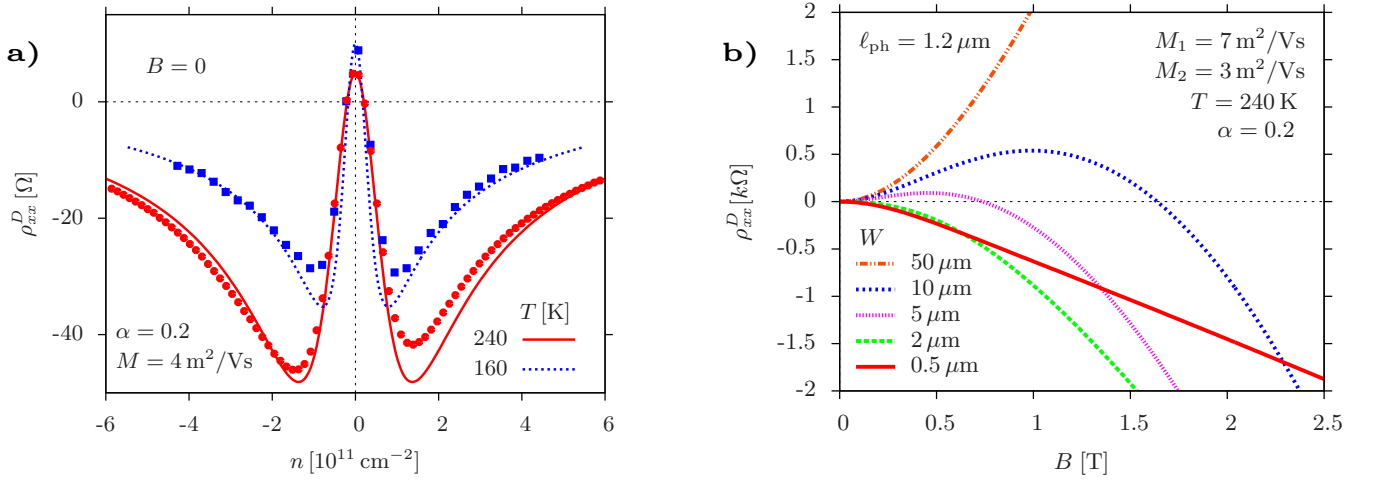}}
\caption{Left Panel: Fit of the zero-field drag resistivity. This fit is used to determine the phenomenological friction coefficient
  $\gamma$. Note, that the theoretical curve in this plot is shifted vertically to fit the shape of the experimental data. The obtained $\gamma$ is subsequently used to calculate magnetodrag. Right Panel: magneto-drag at charge neutrality calculated for different sample widths in the case of inequivalent layers.}
 \label{fig:B}
\end{figure}

\section{Relaxation rates}

In order to determine how the above results depend on carrier density,
we need to estimate the dependence of the scattering rates
$\tau_Q^{-1}$, $\tau_{\rm ph}^{-1}$ and the phenomenological
coefficient $\gamma$ on the chemical potential and temperature. While
comparing our theory to the experimental data, we use the
interpolation formulas listed below, such that the only remaining
fitting parameters are the electron-electron and electron-phonon
coupling constants ($\alpha$ and $g_{\rm ph}$, respectively).

\subsection{Momentum relaxation rate}

Since the drag at zero magnetic field is entirely determined by the
"friction coefficient" $\gamma$ we can use earlier theoretical work \cite{Narozhny12A,Schuett13A}
and actual experimental data to fit
$\gamma$. The result is a function of a single parameter that has the
following general properties: (i) $\gamma\propto\alpha^2$, (ii)
$\gamma\rightarrow 0$ for $\mu\gg T$, and (iii) $\gamma$ remains
finite at the Dirac point.

For $T=240\,$K, the system can be approximately regarded as
ballistic. In this case, microscopic calculations show that at the
Dirac point $\gamma(\mu=0)\sim\alpha^2$, while far away from the Dirac
point $\gamma(\mu\gg T)\sim\alpha^2 T^2/\mu^2$. In the intermediate
region $\mu\sim T$ the momentum relaxation rate is given by a
complicated integral, but effectively it just interpolates between the
two limits. Given that the phenomenological theory is only accurate in
those two limits, we may use a simple interpolation
\be
\gamma = \frac{\alpha^2}{1+\mu^2/2T^2}.
\e

For $T=160$\,K, the system enters the diffusive regime which complicates
the theory. We use our earlier estimates of the drag resistance in the diffusive
regime \cite{Narozhny12A} and fit the dimensionless relaxation rate $\gamma$
to the drag data at zero magnetic field, see Fig.~\ref{fig:B}. Within the phenomenological theory, the zero-field
drag resistance is given by Eq.~(\ref{zfrd}). Using this result and respecting the above general restrictions on $\gamma$ we arrive at the empirical expression, which is applicable for not too high values of the chemical potential:
\be
\gamma\approx\frac{\alpha^2}{\sqrt{4+(\mu/T)^2 (1+0.35\ln^2(\mu/T))}}.
\e

\subsection{Energy relaxation rate}

The lowest order phonon contribution to the electron-hole
recombination in graphene is kinematically forbidden (within the same
valley). There are, however, many possible mechanisms of recombination
\cite{Foster09A,Song12}, all of them involving phonons. Therefore,
we phenomenologically regard the time and length scale of electron and
hole recombination as $\tau_{\textrm{ph}}$ and $\ell_{\textrm{ph}}$.

The microscopic theory \cite{Schuett13A} includes
thermoelectric effects formulated in terms of energy currents. In
graphene the energy current is equal to the total momentum
\cite{Schuett13A}. Therefore, the corresponding relaxation processes do
not require recombination and can be directly attributed to phonons.

The quasiparticle relaxation rate due to the scattering on phonons can
be estimated with the help of the Fermi golden rule. The result is
\be
\frac{1}{\tau_{\rm ph}} = \frac{g_{\rm ph}^2(T) T}{\cosh(\mu/T)}.
\e
This estimate includes the disorder-assisted electron-phonon
scattering processes \cite{Song12} as well as phonon-induced
intervalley scattering and is valid for temperatures below the Debye
temperature in hBN.

\begin{figure}
\centerline{
\includegraphics[width=0.5\columnwidth]{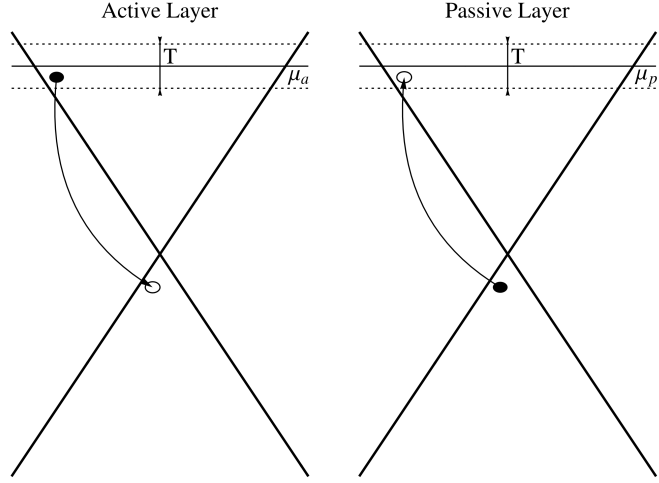}}
\caption{Sketch of the quasi-particle relaxation between two layers}
 \label{fig:C}
\end{figure}

\subsection{Imbalance relaxation}

The quasiparticle recombination rate due to the energy transfer
between the graphene layers can also be estimated by means of the
Fermi Golden Rule in the two limits $\mu\ll T$ and $\mu\gg
T$. Interpolating between the two limits, we obtain the following
estimate

\be
\frac{1}{\tau_Q}=\frac{\alpha^2\sqrt{T^2+\mu^2}}{4\cosh(\mu/T)}.
\e
The corresponding process is illustrated in Fig.~\ref{fig:C}.


\begin{thebibliography}{99}
\bibitem{fn1} The hydrodynamic description of drag in graphene derived
  in Ref.~\onlinecite{Schuett13} was justified by the singular
  behavior of the collision integral due to kinematics of Dirac
  fermions. This singularity leads to the fast unidirectional
  thermalization and allows one to select the relevant eigenmodes of
  the collision integral \cite{Fritz08,Mueller08}. Projecting the
  collision integral onto these modes, one arrives at the effective
  model in terms of energy and electric currents, which is equivalent
  to Eq.~(\ref{Drude0}) with the generalized force (\ref{sim}). At the
  Dirac point the energy current is equivalent to the quasi-particle
  current $\bb{P}$.
\bibitem{fn2} The lowest order phonon contribution to the
  electron-hole recombination in graphene is kinematically
  forbidden. There are, however, many possible mechanisms of
  recombination \cite{Foster09,Levitov12}, all of them involving
  phonons. Therefore, we phenomenologically regard the time and length
  scale of electron and hole recombination as $\ell_{\textrm{ph}}$ and
  $\tau_{\textrm{ph}}$. Moreover, the microscopic theory
  \cite{Schuett13,us2} includes thermoelectric effects formulated in
  terms of energy currents rather. In graphene the energy current is
  equal to the total momentum \cite{Schuett13}. Therefore, the
  corresponding relaxation processes do not require recombination and
  can be directly attributed to phonons. At the same time, the
  hydrodynamic description in terms of the energy currents yields only
  the power-law decay of the magneto-drag at $\mu_i\gg T$, in contrast
  to the exponential collapse shown in Fig.~\ref{fig:results}, panels
  B and E.
\end{thebibliography}
\end{document}